\documentclass[useAMS,usenatbib,useAMS,referee]{mn2e}

\usepackage{times}

\usepackage{graphicx}
\usepackage{amsmath, amssymb}
\usepackage{txfonts}
\usepackage{amssymb}

\usepackage{graphicx,graphics}
\usepackage{natbib}

\title[Scaling law of BHs]{A universal scaling law of black hole activity including gamma-ray bursts}
\author[Wang \& Dai]{F. Y. Wang$^{1,2}$\thanks{E-mail:
fayinwang@nju.edu.cn} \& Z. G. Dai$^{1,2}$\\
$^1$School of Astronomy and Space Science, Nanjing University,
Nanjing 210093, China\\
$^2$Key Laboratory of Modern Astronomy and Astrophysics (Nanjing
University), Ministry of Education, Nanjing 210093, China}
\begin{document}

\date{}

\pagerange{\pageref{firstpage}--\pageref{lastpage}} \pubyear{}

\maketitle

\label{firstpage}

\begin{abstract}

Previous works show that a correlation among radio luminosity, X-ray
luminosity, and black hole (BH) mass from stellar-mass BHs in X-ray
binaries to supermassive BHs in active galactic nuclei (AGNs), which
leads to the so-called fundamental plane of BH activity. However,
there are two competing explanations for this fundamental plane,
including the jet-dominated model and the disk-jet model. Thus, the
physical origin of this fundamental plane remains unknown. In this
paper, we show that the X-ray luminosities, radio luminosities and
BH masses of gamma-ray bursts (GRBs) and M82 X-1 also show a similar
distribution. The universal scaling law among stellar-mass,
intermediate and supermassive BH systems, together with the fact
that radio and X-ray emission of GRBs originates from relativistic
jets, reveals that the fundamental plane of BH activity is
controlled by a jet, i.e., the radio and X-ray emission is mainly
from the jet. Our work also suggests that the jets are
scale-invariant with respect to the BH mass.

\end{abstract}


\section{Introduction}

A long-standing and intriguing question is that the similarity among
stellar, intermediate and supermassive black holes (BHs) and their
relativistic jets \citep{White84,Mir99,Meier03}. Some interesting
results have been found, including the similar accretion process
\citep{McHardy06}, radiation mechanism \citep{Wang14}, statistical
properties of X-ray flares \citep{WangF14} and energetics of
relativistic jets \citep{Nemmen12}.

A tight correlation between the radio luminosity and X-ray
luminosity has been firstly discovered in X-ray binaries GX 339-4
and V404 Cyg \citep{Corbel00,Gallo03}. After considering the BH
mass, AGNs show a similar correlation, which is called the
fundamental plane of BH activity \citep{Merloni03,Falcke04}. This
correlation has been confirmed by later studies
\citep{Kording06,Li08,Yuan09,Gultekin14}. \cite{Lu15} showed that
radio luminosity correlated with X-ray luminosity for gamma-ray
bursts (GRBs). However, the power-law slope is slightly different
with the one found in X-ray binaries \citep{Corbel00,Gallo03}.
Although some doubts on the existence of the fundamental plane have
been raised, Merloni et al. (2006) presented that the fundamental
plane correlation cannot be a distance artifact, and that it must
represent the intrinsic characteristic of BHs. The fundamental plane
attempts to unify sources associated with BHs, over a large range of
masses and luminosities, from Galactic sources to AGNs. However,
some XRBs might not follow the fundamental plane \citep{Corbel04}.
Some other outliers have been found
\citep{Jonker10,Coriat11,Ratti12}. For the sub-Eddington objects,
i.e., low/hard state of XRBs and low-luminosity AGNs, K\"{o}rding et
al. (2006) found that the fundamental plane is much tighter than the
full sample. Therefore, they concluded that the fundamental plane
should be most suitable for radiatively inefficient BH sources.
Another form of the fundamental plane for radiatively efficient BHs
is established by \cite{Dong14} and \cite{Xie17}.

However, the physical origin of the fundamental plane and its
relationship with the physical properties of the sources are
mysterious \citep{Narayan05}. There are at least two possible
explanations for the origin of the fundamental plane. The first
explanation is that the radio and X-ray emission for BH sources is
attributed to synchrotron emission from a jet in the jet-dominated
state. Because the radio and X-rays emission is powered by the jet,
a tight correlation between them could be expected
\citep{Falcke04,Yuan05}. For example, \cite{Markoff03} modeled the
broadband spectrum of X-ray binary GX 339-4 and explained the
fundamental plane using the jet model. Although, the disk component
may contribute $\leq 20$ per cent in the soft X-ray band for
low/hard state of some black hole X-ray binaries
\citep{Poutanen98,Fender01,Markoff04,Homan05,Markoff05,Remillard06}.
On the other hand, the X-ray emission may originate from the
accretion flow, while the radio emission is produced by the
relativistic jet. Meanwhile, the accretion flow and the jet are
strongly coupled, so that the radio and X-ray luminosity is
correlated \citep{Merloni03,Heinz03}. When the X-ray luminosity is
below a critical value, the X-ray radiation should be dominated by
emission from the jet rather than from the accretion flow
\citep{Yuan05}. \cite{Yuan09} showed that the X-ray emissions from
IC 1459, M32, M81, M84, M87, NGC 3998, NGC 4594, NGC 4621, and NGC
4697 are jet-dominated. \cite{Younes12} fitted the spectral energy
distributions of six LINERs, and showed that the X-ray emission is
jet-dominated. Hence, a further clarification is urgently needed.
Moreover, whether intermediate-mass BHs follow this correlation
remains unknown. Some candidates of intermediate-mass BHs have been
found, such as M82 X-1 \citep{Pasham14}, and HLX-1
\citep{Farrell09}.

Gamma-ray bursts (GRBs) are the most powerful electromagnetic
explosions in the universe. The progenitors of GRBs are thought to
be massive stars \citep{Woosley93} or mergers of compact stars
\citep{Eichler89,Nakar07}. So the central engine of GRBs may be
stellar-mass BHs. The radiation in X-ray and radio bands is well
understood, which originates from relativistic jets
\citep{Zhang04,Piran04,Meszaros06,Gehrels09,Kumar15}. GRBs, as
bright lighthouses in the distant universe, are ideal tools to probe
the properties of the early universe \citep{Wang15}. There are some
similar properties between GRBs and AGNs, such as the variability
property \citep{Wu15}, and radiation mechanism \citep{Lyu14}.
However, whether GRBs follow the fundamental plane is unknown.

In this paper, we investigate the fundamental plane of BH activity
including GRBs and the intermediate-mass BH M82 X-1. This paper is
organized as follows. In Section 2, we present the sample of GRBs.
The fitting results are presented in Section 3. Section 4 gives
summary.

\section{Samples}

We collect the data of GRBs from previous literature. Because the
luminosity of GRBs is required, the GRBs with redshift measurements
are selected. Meanwhile, the radio and X-ray observations are also
required. There are 31 GRBs in our sample, which are listed in Table
1. In this table, columns 2 is the redshift of a GRB. The prompt
emission flux measured between energy range of columns 5 and 6 is
listed in columns 3. The peak energy $E_{\rm peak}$ of the spectrum
is in columns 4. The low-energy and high-energy power-law spectral
indices $\alpha$ and $\beta$ are given in columns 7 and 8,
respectively. When the values of $\alpha$ and $\beta$ are unknown,
the typical values $\alpha=1.1\pm 0,4$ and $\beta=2.2\pm0.4$ are
adopted \citep{Schaefer07}. The radio flux at observed frequency
(column 10) is presented in Column 9. Column 11 is the references of
parameters. For the radio flux, we use the peak radio flux from
Chandra \& Frail (2012). The isotropic radio luminosity at 5 GHz is
calculated by \citep{Chandra12}
\begin{equation}
L_R=4\pi d_L^2\nu F_R F_{\rm beam} (1+z)^{-\alpha_1-1},
\end{equation}
where $d_L$ is the luminosity distance, $F_R$ is radio flux at 5
GHz, and $\alpha_1=1/3$ is the spectral slope in the slow cooling
regime \citep{Sari98}. The collimation-corrected luminosity is given
by $L_{R,c}=L_RF_{\rm beam}$, where $F_{\rm beam}=1-\cos\theta_j$ is
the beaming factor, and the jet opening angle $\theta_j$ is adopted
from Wang et al. (2014). The observed radio flux is converted to the
flux at 5 GHz using spectral slope $\alpha_1=1/3$. Because the X-ray
flux declines quickly after the prompt emission, we adopt the X-ray
flux at the prompt emission as an approximation of the peak value.
The X-ray luminosity in $2-10$ keV is $L = 4\pi d^2_{L}P_{\rm
bolo}F_{\rm beam}$, where $P_{\rm bolo}$ can be derived by
\begin{equation}
P_{\rm bolo} = P  \ {\times} \ \frac{\int_{2/(1 + z)}^{10/(1 + z)}{E
\Phi(E) dE}} {\int_{E_{\rm min}}^{E_{\rm max}}{E\Phi(E) dE}},
\label{eq: defpbolo}
\end{equation}
where $P$ in units of erg cm$^{-2}$ s$^{-1}$ is the observed peak
flux between $E_{\rm min}$ and $E_{\rm max}$, and $\Phi(E)$ is the
Band function \citep{Band93}. If the flux $P$ is in units of photons
cm$^{-2}$ s$^{-1}$, the $P_{\rm bolo}$ is
\begin{equation}
P_{\rm bolo} = P  \ {\times} \ \frac{\int_{2/(1 + z)}^{10/(1 + z)}{E
\Phi(E) dE}} {\int_{E_{\rm min}}^{E_{\rm max}}{\Phi(E) dE}}.
\label{eq: defpbolo}
\end{equation}
The mass of a GRB progenitor is still uncertain. The theoretical
model of long GRBs is thought to be collapse of a massive star to a
stellar-mass BH \citep{Woosley93}. While mergers of neutron
star-neutron star or neutron star-BH binaries may produce short
GRBs. So we adopt two assumptions for the mass, $3M_\odot$ and
$10M_\odot$, and 10\% relative uncertainties. These values are
consistent with numerical simulations \citep{MacFadyen99,Dessart12}.

Generally, the X-ray emission of GRB prompt emission is from
internal shock, which is formed by collisions of relativistic shells
\citep{Rees94}. After collisions, the shells will combine to one
shell. This shell sweeps the external medium and produces the
external shock. The radio emission of GRBs is from external shock,
which is generated by the relativistic shell from central engine
sweeping the external medium \citep{Sari98}. So this jet must carry
the memory of the central engine. It is reasonable to make an
analogy between AGN and GRBs. For example, \cite{Nemmen12} showed
that jets produced by AGNs and GRBs exhibit the same correlation
between the kinetic power and the gamma-ray luminosity.

For the X-ray binary (XRB) and AGN sample, we use the sample from
K\"{o}rding et al. (2006). This sample includes XRBs in the low/hard
state. There are also low-luminosity AGN (LLAGN), LINERs (low
ionization nuclear emission region), FR I Radio Galaxies and BL Lac
objects in this sample. The masses of supermassive BHs are from
$10^7M_\odot$ to $10^9M_\odot$. Because of the existence of
different accretion states for XRBs, appropriate sources must be
selected when studying the fundamental plane
\citep{Yuan05,Ho05,Kording14}. Radio observations confirmed that the
low/hard state is associated with a jet \citep{Fender01}. Different
states of XRBs may follow different radio-X-ray correlations
\citep{Yuan05,Kording06}. We include GX 339-4 with $8.0\pm
2.0~M_\odot$ \citep{Kording06}, V404 Cyg with $12.0\pm 2.5~M_\odot$
\citep{Kording06}, 4U 1543-475 with $9.4\pm 1.0~M_\odot$
\citep{Remillard06}, and XTE 1118+480 with $6.8\pm 0.5~M_\odot$
\citep{Remillard06}. For Sgr A$^*$, we include the hard X-ray flare
observed by \cite{Baganoff01}, as the flare may be due to jet
emission \citep{Markoff01}. Besides the flare period, the quiescent
Sgr A$^*$ deviates from the fundamental plane
\citep{Falcke04,Kording06}. Recent study shows that the mass of M82
X-1 is $428\pm105~M_\odot$, which belongs to an intermediate-mass BH
\citep{Pasham14}. We must note that the mass is determined by
extrapolating the inverse-mass scaling that holds for stellar-mass
black holes \citep{Pasham14}, but whether this scaling can be
extrapolated is unclear. The simultaneous observation shows that the
X-ray luminosity is about $2.4\times10^{40}$ erg s$^{-1}$ and radio
luminosity is $(6.7\pm1.3)\times10^{40}$ erg s$^{-1}$
\citep{Kaaret06}.

It must be noted that the XRBs in the low/hard state are not
flaring, and the Sgr A$^*$ is flaring in our sample. There are two
states for XRBs: the high/soft state with a soft power-law spectrum
dominated by a thermal ``bump'', and the low/hard state
characterized by a dominant hard power-law and weak-to-absent
thermal spectrum. On the other hand, some low-power AGN classes seem
to lack evidence of a blue bump, and are analogous to the hard-state
XRBs. These are FR I radio galaxies, BL Lacs and LINERs. The Sgr
A$^*$ may also be in this category. Interestingly, the flares in Sgr
A$^*$ (Baganoff et al. 2001) have a hard spectrum, so it may be
analogous to the low/hard state in XRBs. Observations showed that
XRBs produce powerful collimated outflows in the low/hard state
\citep{Fender01}. The evidence of jet from Sgr A$^*$ in flaring
state also was found using high-resolution Very Large Array images
and ultra-deep imaging-spectroscopic data from Chandra X-ray
Observatory \citep{Li2013}. In quiescent state, independent
evidences showed that a jet does not appear in Sgr A$^*$
\citep{Shen05,Doeleman08}. So our sample contains the non-flaring
XRBs in the low/hard state and the flaring Sgr A$^*$.

Although, the radiation in an advection-dominated accretion flow
(ADAF) \citep{Narayan94} is dominated by thermal Comptonization, in
agreement with observations of XRBs in the low/hard state
\citep{Narayan05}. For example, combining the thin disk model and
the ADAF model, \cite{Narayan96} and \cite{Esin97} showed that the
various spectral states of XRBs could be understood. The ADAF model
or its variants \citep{Blandford99,Narayan00}, can explain many
observations of XRBs \citep{Esin98,Esin01}. Meanwhile, as discussed
above, the X-ray emission of the low/hard state XRBs can also be
produced by a jet. When the X-ray luminosity is below a critical
value (10$^{-5}$ to 10$^{-6}$ $L_{\rm edd}$), the X-ray emission
should be dominated by the emission from the jet \citep{Yuan05}. So
in this paper, we assume that the X-ray emissions of low/hard state
XRBs are mainly from jets.

The X-ray luminosity and radio luminosity used in fundamental plane
should be taken quasi-simultaneously. But the simultaneous
observations are rare. For AGNs, the radio and X-ray observations
are usually non-simultaneous. Sometimes, there is more than a year
between the different observations \citep{Kording06}. But the
orientation of this uncertainty is likely to be isotropic
\citep{Kording06}. \cite{Bell11} showed that the time-lag between
the X-ray and radio bands is about 40 days for NGC 7213.
Interestingly, NGC 7213 lies very close to the best-fitting
correlation of the fundamental plane. \cite{Miller10} proposed that
an individual source might not be rigidly governed by the
fundamental plane on short timescales. But it follows the
fundamental plane in a time-averaged sense. Although the emission
regions producing the radio and the X-ray emission may be different.
For GRBs, because of lacking the radio emission in the prompt phase,
we use the peak luminosities of X-ray and radio bands, which is
similar to the average radio and X-ray luminosities for AGNs.
Meanwhile, the time lags between X-ray and radio observations for
GRBs are from 1 day to 50 days \citep{Chandra12}. So the time lags
is comparable to that of AGNs.

\section{Results}

The fundamental plane reads
\begin{equation}
\log L_X=a_R\log L_R+a_M\log M+b,
\end{equation}
where $L_X$ is the X-ray luminosity, $L_R$ is the radio luminosity
at $5$ GHz, $M$ is the mass of BH, and $a_R$, $a_M$ and $b$ are
fitting parameters. In order to derive the best-fitting parameters
for three variables with errors, we use the merit function
\citep{Press92,Merloni03,Kording06}, which is defined as:
\begin{equation}
\hat{\chi}^2 = \sum_{\text i} \frac{(y_{\text i} - b - \sum_{\text
j} a_{\text j} x_{\text ij})^2 }{\sigma^2_{\text y_i} + \sum_{\text
j} (a_{\text j} \sigma_{\text x_{\text ij}})^2},
\end{equation}
where $y_i$ denotes the X-ray luminosity, $x_{1j}$ is the radio
luminosity, $x_{2j}$ is the mass of BH, and $\sigma$ are the
corresponding uncertainties. Because the standard $\chi^2$ fit only
considers the scatter of variable in the $y$-axis. This merit
function includes all the errors of parameters.

Figure 1 shows the fundamental plane of BH activity including the
XRBs and AGNs from K\"{o}rding et al. (2006), the intermediate-mass
BH M82 X-1, and GRBs. In this figure, the mass of BH for GRB is
fixed to be $3M_\odot$. The value of $y$ axis is $\log L'_X=\log
L_X-a_M\log M$. Using the merit function, we derive the best-fitting
results for the sample of K\"{o}rding et al. (2006) are
$a_R=1.41\pm0.12$, $a_M=-0.87\pm0.17$, and $b=-4.90\pm3.35$. While
for the GRB sample, the parameters are $a_R=1.38\pm0.17$,
$a_M=-0.87\pm0.20$, and $b=-5.02\pm3.20$. The solid blue line is the
best fit of the combined sample. The parameters are
$a_R=1.40\pm0.18$, $a_M=-0.89\pm0.22$, and $b=-5.0\pm3.40$. The
fitted result of the jet-dominated model is $a_R=1.38$, $a_M=-0.81$
\citep{Falcke04}, while the fitted result of disk-jet model is
$a_R=1.64$, and $a_M=-1.3$ \citep{Merloni03}. Our results are
dramatically consistent with the jet-dominated model. Interestingly,
the M82 X-1 with about $400M_\odot$ also follows this correlation.
Figure 2 presents the fundamental plane of black hole activities
same as Figure 1, except that the mass of BH for GRB is assumed as
10$M_\odot$. The fitting parameters of GRBs are $a_R=1.41\pm0.15$,
$a_M=-0.89\pm0.20$, and $b=-5.50\pm3.60$, which is shown as the
black line. The blue line shows the best fit of the whole sample.
The intermediate-mass BH M82 X-1 also follows this correlation.

The X-ray and radio emission of GRBs is highly beamed
\citep{Rhoads97}. So the intrinsic luminosity should be calculated
by $L=L_{\rm iso}F_{\rm beam}=L_{\rm iso}(1-\cos \theta_j)$, where
$L_{\rm iso}$ is the isotropic luminosity and $\theta_j$ is the jet
opening angle. For AGNs, $F_{\rm beam}=1-\cos 1/\Gamma$, where
$\Gamma$ is the bulk Lorentz factor of the outflow, because
$\theta_j<1/\Gamma$ for AGNs. The jet opening angle $\theta_j$ for
GRBs can be derived from the break time of the afterglow light
curve. We use the value of $\theta_j$ from Wang et al. (2011) and
Wang et al. (2014). For AGNs, the Lorentz factor and viewing angle
can be estimated from their variable brightness and apparent jet
speed \citep{Ghisellini93}. Because of lack of observations, it's
very hard to derive all the Lorentz factor of AGNs. Moreover, if the
X-rays originate from the disk or corona, they will not be beamed.
For jet models, the X-ray emission may be beamed like the radio
emission or have a different beaming patterns. So we can only assume
isotropy \citep{Kording06}. In Figure 3, we show the fundamental
plane of BH activity after correcting the beaming factor of GRBs.
The masses of GRBs are assumed as $3M_\odot$. The fitting result of
GRBs is shown as the black line with parameters, $a_R=1.375\pm0.19$,
$a_M=-0.88\pm0.22$ and $b=-4.35\pm3.10$. These results are
consistent with that of XRBs, M82 X-1 and AGNs. This consistence can
be expected. As in the jet-dominated model, the X-ray and radio
luminosities are both corrected by a factor of $(1-\cos \theta_j)$.
By considering the relativistic beaming, Plotkin et al. (2012) found
that relativistically beamed BL Lac objects fit well onto the
fundamental plane. The slopes of the fundamental plane are almost
unchanged when the beaming effect is included \citep{Plotkin12}. The
fundamental plane is weakly dependent on the beaming factor
\citep{Kording06}.

\section{Summary}
In this paper, we have found that a correlation exists between the
radio luminosity, X-ray luminosity, and BH mass among stellar-mass,
intermediate-mass, and supermassive BHs, which is called the
fundamental plane of BH activity. The analogy known to exist between
XRBs in the low/hard state and AGNs can be extended to GRBs and
intermediate-mass BH M82 X-1. Our results suggest that the
underlying physics of the fundamental plane is that the X-ray and
radio emission is produced by a jet, which is called the
jet-dominated model. The jets are scale-invariant with respect to
the BH mass. So it is possible to use the fundamental plane to
estimate BH masses, especially for the intermediate-mass BHs and
GRBs.

\section*{Acknowledgements}
We thank the anonymous referee for constructive comments. We thank
F. Yuan, Y. C. Zou and B. Zhang for helpful discussions. This work
is supported by the National Basic Research Program of China (973
Program, grant No. 2014CB845800) and the National Natural Science
Foundation of China (grants 11422325, 11373022, and 11573014), and
the Excellent Youth Foundation of Jiangsu Province (BK20140016).

\clearpage
\begin{figure}
\centering
\includegraphics[width=1.0\textwidth]{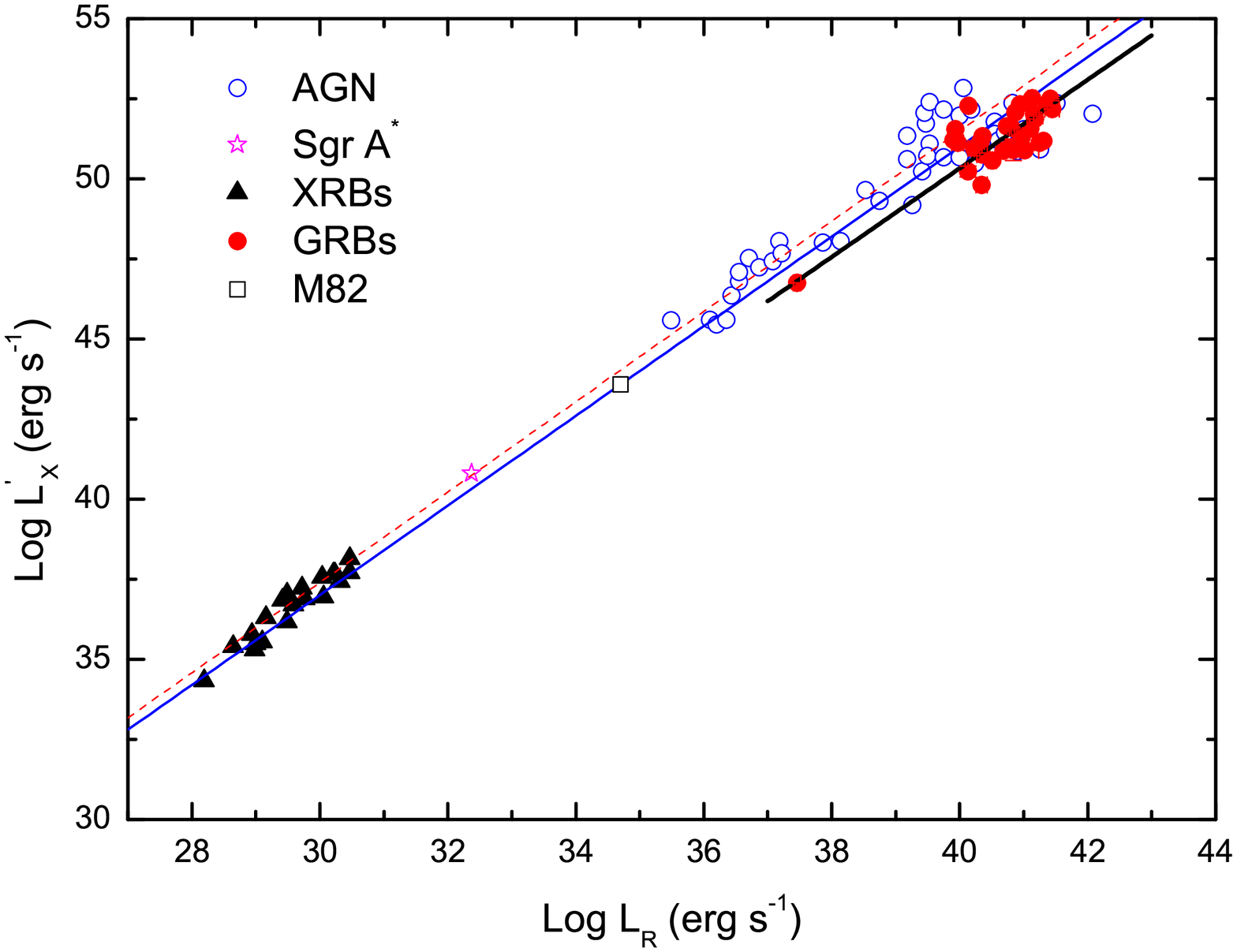}
\caption{\label{Fig1} The fundamental plane of BH activity. The mass
of BH for GRB is assumed to be 3$M_\odot$. The black line represents
the best fitting of GRBs. The red line is the best fitting for the
XRBs and AGNs, and the blue line is the best fitting for the whole
sample ($\log L_X=(1.41\pm0.12)\log L_R-(0.87\pm0.17)\log
M-(4.90\pm3.35)$). The square is M82 X-1, which is dramatically
consistent with the best fitting. So the universal scaling law of BH
activity exists in stellar-mass, intermediate and supermassive BHs.}
\end{figure}

\clearpage
\begin{figure}
\centering
\includegraphics[width=1.0\textwidth]{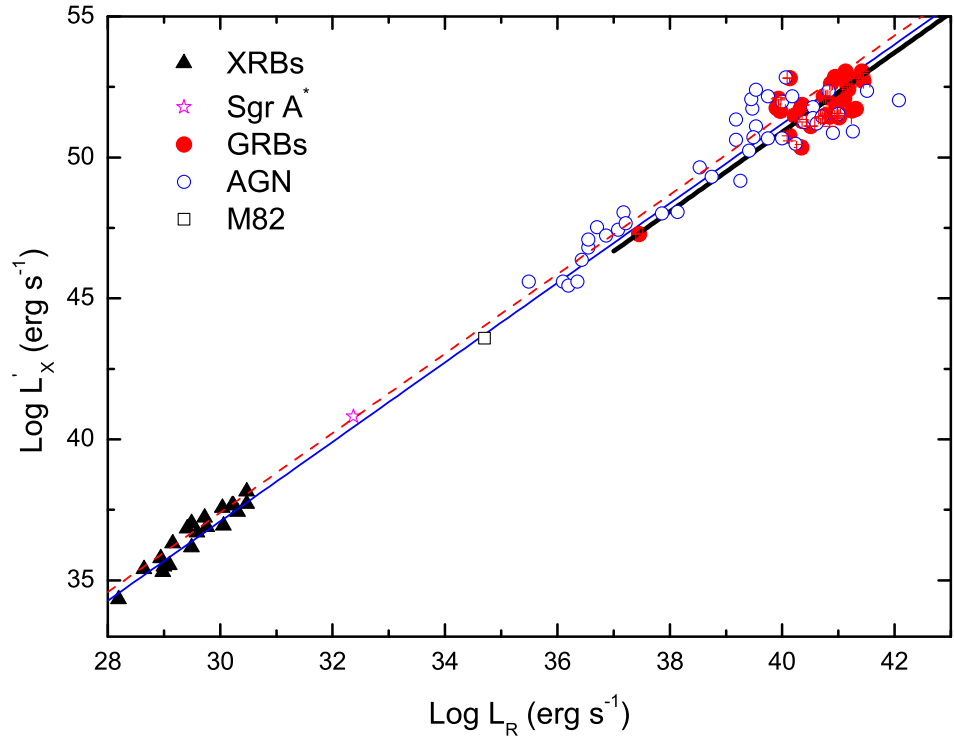}
\caption{\label{Fig2} The same as Figure 1, except that the masses
of GRB BHs are fixed as 10$M_\odot$. The best fitting results for
the whole sample is $\log L_X=(1.41\pm0.20)\log
L_R-(0.88\pm0.18)\log M-(5.21\pm3.50)$. }
\end{figure}

\clearpage
\begin{figure}
\centering
\includegraphics[width=1.0\textwidth]{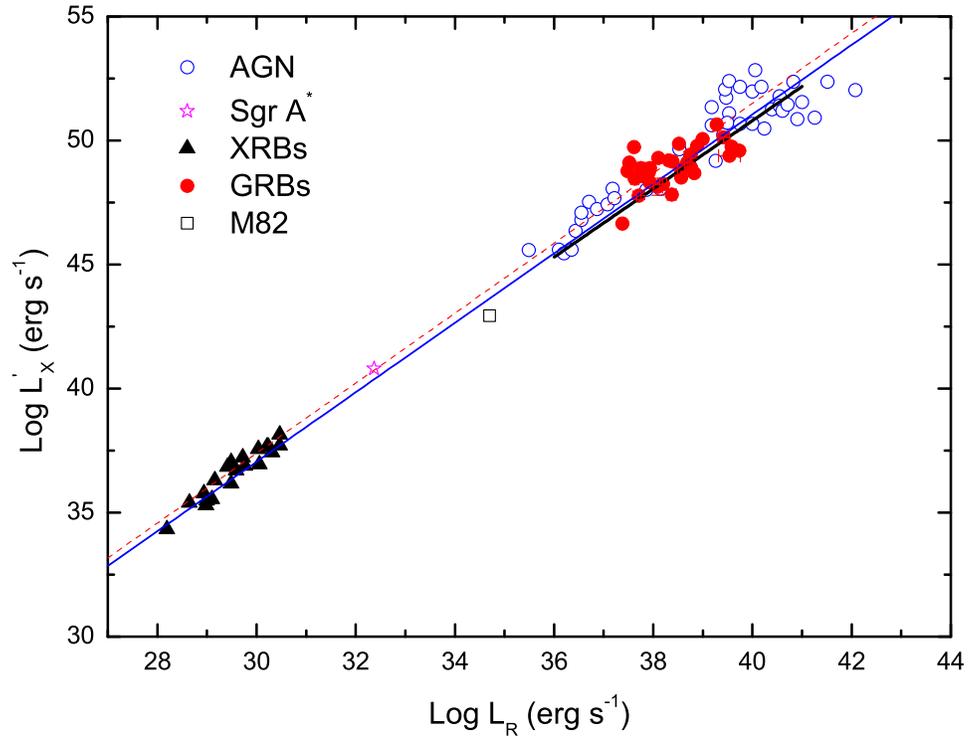}
\caption{\label{Fig3} The fundamental plane of BH activity. The
black line represents the best fitting of GRBs with beaming
corrected. The mass of BH for GRB is assumed to be 3$M_\odot$. The
red line is the best fitting for the XRBs and AGNs from K\"{o}rding
et al. (2006), and the blue line is the best fitting for the whole
sample. The square represents the M82 X-1, which is dramatically
consistent with the best fitting. So the universal scaling law of BH
activity is not affected by the beaming effect.}
\end{figure}
\clearpage

\begin{table}
\centering
\begin{tabular}{ccccccccccc}
\hline \hline
GRB & $z$ &$P^a$& $E_{peak}$& $E_{min}$ &  $E_{max}$ & $\alpha$ & $\beta$ & $F_{R}$& Frequency & Ref$^{b}$\\
    &     &     &  keV  & keV & keV &   &  & $\mu$Jy  & GHz &   \\
\hline
970508  &   0.835   &   7.4E-7$\pm$7E-8    &   389$\pm$40 &   40  &   700 &   -1.19$\pm$0.1   &   -1.83$\pm$0.4    &   780$\pm$13 &   4.86 &  1,2 \\
970828  &   0.958   &   3.0E-6$\pm$3E-7   &   298$\pm$30 &   50  &   300 &   -0.7$\pm$0.1    &   -2.07$\pm$0.4   &   144$\pm$31 &   8.46   &  1,2 \\
980703  &   0.966   &   2.4$\pm$0.06 &   254$\pm$25 &   50  &   300 &   -1.31$\pm$0.1   &   -2.4$\pm$0.4    &   1055$\pm$30    &   4.86   &  1,2 \\
990510  &   1.619   &   8.17$\pm$0.08    &   126$\pm$10 &   50  &   300 &   -1.28$\pm$0.1   &   -2.67$\pm$0.4   &   255$\pm$34 &   8.46   &  1,2 \\
010222   &   1.477   &   8.6E-6$\pm$2E-7 &   309$\pm$12 &   40  &   700 &   -1.35$\pm$0.19   &   -1.64$\pm$0.02   &   93$\pm$25  &   8.46   &  1,2 \\
021004   &   2.33    &   0.89$\pm$0.20    &   80$\pm$38  &   30  &   400 &   -1.01$\pm$0.18   &   -2.2$\pm$0.4    &   470$\pm$26 &   4.86  &  1,2  \\
030226   &   1.986   &   0.99$\pm$0.17    &   97$\pm$17  &   30  &   400 &   -0.89$\pm$0.16   &   -2.2$\pm$0.4    &   171$\pm$23 &   8.46  &  1,2 \\
030329   &   0.169   &   72.2$\pm$3.8    &  68$\pm$2.2  &   30  &   400 &   -1.26$\pm$0.02   &   -2.28$\pm$0.06 & 10337$\pm$33   &   4.86  &  1,2  \\
050416A &   0.65    &   4.8$\pm$0.4 &   15$\pm$2.7  &   15  &   350 &   -1.1$\pm$0.4    &   -3.4$\pm$0.4  &   485$\pm$36 &   4.86   &  1,2 \\
050603   &   2.821   &   31.8$\pm$1.7    &   344$\pm$52 &   15  &   350 &   -1.03$\pm$0.06   &   -2.03$\pm$0.1   &   377$\pm$53 &   8.46  &  1,2  \\
050820A &   2.615   &   1.3$\pm$0.2 &   246$\pm$40 &   15  &   150 &   -1.25$\pm$0.10   &   -2.2$\pm$0.4    &   150$\pm$31 &   8.46   &  1,2 \\
050904   &   6.29    &   0.8$\pm$0.1 &   436$\pm$90 &   15  &   150 &   -1.11$\pm$0.06   &   -2.2$\pm$0.4    &   76$\pm$14  &   8.46  &  1,2  \\
051022   &   0.809   &   1.0E-5$\pm$8E-7  &  510$\pm$20 &   20  &   2000    &   -1.176$\pm$0.02  &   -2.2$\pm$0.4  & 268$\pm$32 &   8.46  &  1,2  \\
060218   &   0.033   &   0.021$\pm$0.005   &   4.9$\pm$0.3 &   15  &   150 &   -0.84$\pm$0.1   &   -2.2$\pm$0.4    &   245$\pm$50 &   4.86   &  2,3 \\
070125   &   1.548   &   2.25E-5$\pm$3.4E-6    &   367$\pm$51 &   20  & 10000   &   -1.1$\pm$0.1 &   -2.08$\pm$0.15 & 1028$\pm$16 &   8.46  &  2,4  \\
071003 &1.604  &   1.11E-6$\pm$1.0E-7    &   799$\pm$100 &   15  &   350 &   -1.31$\pm$0.07   &   -2.2$\pm$0.4    &   616$\pm$57 &   8.46   &  2,3 \\
071010B    &   0.947   &   5.48E-7$\pm$5E-8   &  52$\pm$6  &   15  &   150 &   -1.5$\pm$0.2   &   -2.2$\pm$0.4   &   227$\pm$114 &   4.86  &  2,5  \\
090323   &   3.57    &   5.96E-6$\pm$1.06E-6    &   416$\pm$76 &   20  &   10000   &   -0.96$\pm$0.09   &   -2.09$\pm$0.22   &   243$\pm$13 &   8.46  &  2,6  \\
090328   &   0.736   &   12.2E-5$\pm$2.5E-6    &   592$\pm$237 &   20  &   8000    &   -1.04$\pm$0.1   &   -2.05$\pm$0.9   &   686$\pm$26 &   8.46   &  2,7 \\
090423   &   8.26    &   1.02E-7$\pm$8E-8   &   49$\pm$3.8  &   15  &   150 &   -0.77$\pm$0.08   &   -2.2$\pm$0.4    &   50$\pm$11  &   8.46  &  2,5,8  \\
090424   &   0.544   &   9.12E-6$\pm$1.4E-7    &   162$\pm$3.4 &   8   &   35000   &   -1.02 $\pm$0.01  &   -3.26$\pm$0.14   &   236$\pm$37 &   8.46   &  2,9 \\
090715B  &   3   &  9E-7$\pm$9E-8 &   134$\pm$40 &   20  &   2000    &   -1.1$\pm$0.37    &   -2.2$\pm$0.4    &   191$\pm$36 &   8.46   &  2,9 \\
091020   &   1.71    &   1.88E-6$\pm$2.6E-7    &   187$\pm$34 &   8   &   35000   &   -1.2$\pm$0.06    &   -2.29$\pm$0.18    &   399$\pm$21 &   8.46  &  2,9  \\
100418A  &   0.62    &   6.62E-8$\pm$3.6E-8    &   25$\pm$3.0  &   15  &   150 &   -1.0$\pm$0.1  &   -2.06$\pm$0.3   &   522$\pm$83 &   4.86  &  2,10  \\
100814A  &   1.44    &   7.5E-7$\pm$2.5E-7    &   128$\pm$23 &   20  &   2000    &   -1.1$\pm$0.2   &   -2.2$\pm$0.4   &   496$\pm$24 &   4.5  &  1 \\
991216  &   1.02    &   67.52$\pm$0.23   &   318$\pm$30 &   50  &   300 &   -1.23$\pm$0.1   &   -2.18$\pm$0.4   &   126 &   4.86  &  1,2  \\
000210  &   0.85    &   29.9$\pm$3.0    &   408$\pm$14 &   50  &   300 &   -1.1$\pm$0.4    &   -2.2$\pm$0.4    &   93  &   8.46  &  1,2  \\
050401  &   2.898   &   12.6$\pm$1    &   118$\pm$18 &   20  &   2000    &   -0.9$\pm$0.3  &   -2.55$\pm$0.3   &   122 &   8.46  &  1,2  \\
050525A &   0.606   &   48.0$\pm$0.6  &   81.2$\pm$1.4  &   15  &   350 &   -1.01$\pm$0.06   &   -3.26$\pm$0.2  &   164 &   8.46   &  1,2 \\
050824   &   0.83    &   0.5$\pm$0.2 &   15$\pm$2  &   15  &   150 &   -1.1$\pm$0.4    &   -3.3$\pm$1.0    &   152 &   8.46  &  1,2  \\
060418   &   1.49    &   6.7$\pm$0.2 &   230$\pm$20 &   15  &   150 &   -1.5$\pm$0.1    &   -2.2$\pm$0.4    &   216 &   8.46   &  1,2 \\
071020   &   2.146   &   6.04E-6$\pm$2.08E-6    &   322$\pm$65 &   20  &   2000    &   -0.65$\pm$0.29   &   -2.0$\pm$0.4  &   141 &   8.46  &  2,9  \\

\hline

\end{tabular}
\caption{GRB data.$^a$The reported $P$ value is either in units of
photons cm$^{-2}$ s$^{-1}$ for values larger than 0.01 (not in
scientific notation) or in units of erg cm$^{-2}$ s$^{-1}$ for
values smaller than 0.01 (those in scientific notation). $^b$(1)
Schaefer (2007); (2) Chandra \& Frail (2012); (3) Butler et al.
(2007); (4) Golenetskii et al. (2007); (5) Butler et al. (2010); (6)
Golenetskii et al. (2009a); (7) Golenetskii et al. (2009b); (8) Wang
et al. (2011); (9) Nava et al. (2012); (10)
http://butler.lab.asu.edu/swift. }
\end{table}

\end{document}